\documentclass[preprint,12pt]{elsarticle}

\usepackage{amssymb, amsmath}
\usepackage{graphicx}% Include figure files
\usepackage{hyperref}

\newcommand{\mycode}{CosmoTransitions}

\newcounter{bla}

\journal{Computer Physics Communications}

\begin{document}

\begin{frontmatter}

\title{CosmoTransitions: Computing Cosmological Phase Transition Temperatures and Bubble Profiles with Multiple Fields}

\author{Carroll L. Wainwright}
\address{Department of Physics, University of California, 1156 High St., Santa Cruz, CA 95064, USA}
\ead{cwainwri@ucsc.edu} 

%\date{\today}

\begin{abstract}
\noindent 
I present a numerical package (\mycode) for analyzing finite-temperature cosmological phase transitions driven by single or multiple scalar fields. The package analyzes the different vacua of a theory to determine their critical temperatures (where the vacuum energy levels are degenerate), their super-cooling temperatures, and the bubble wall profiles which separate the phases and describe their tunneling dynamics. I introduce a new method of path deformation to find the profiles of both thin- and thick-walled bubbles. \mycode\ is freely available for public use.

\end{abstract}


\end{frontmatter}

% \linenumbers

% Computer program descriptions should contain the following
% PROGRAM SUMMARY.

{\bf PROGRAM SUMMARY}
  %Delete as appropriate.

\begin{small}
\noindent
{\em Manuscript Title:}  CosmoTransitions: Computing Cosmological Phase Transition Temperatures and Bubble Profiles with Multiple Fields                                     \\
{\em Authors:} Carroll L. Wainwright                                   \\
{\em Program Title:}            
CosmoTransitions                              \\
%{\em Journal Reference:}                                      \\
  %Leave blank, supplied by Elsevier.
%{\em Catalogue identifier:}                                   \\
  %Leave blank, supplied by Elsevier.
{\em Licensing provisions:}    none                               \\
  %enter "none" if CPC non-profit use license is sufficient.
{\em Programming language:} 
Python                                  \\
{\em Computer:}  Developed on a 2009 MacBook Pro. No computer-specific optimization was performed.                                             \\
  %Computer(s) for which program has been designed.
{\em Operating system:}                                       
  Designed and tested on Mac OS X 10.6.8. Compatible with any OS with Python installed.		\\
{\em RAM:} Approximately 50 MB, mostly for loading plotting packages.            \\
  %RAM in bytes required to execute program with typical data.
  % Please give some freely chosen keywords that we can use in a
  % cumulative keyword index.
{\em Classification:}       1.9, 11.1                                  \\
  %Classify using CPC Program Library Subject Index, see (
  % http://cpc.cs.qub.ac.uk/subjectIndex/SUBJECT_index.html)
  %e.g. 4.4 Feynman diagrams, 5 Computer Algebra.
{\em External routines/libraries:}          
  SciPy, NumPy, matplotLib		\\
{\em Nature of problem:}
  %Describe the nature of the problem here.
  I describe a program to analyze early-Universe finite-temperature phase transitions with multiple scalar fields. The goal is to analyze the phase structure of an input theory, determine the amount of supercooling at each phase transition, and find the bubble-wall profiles of the nucleated bubbles that drive the transitions.
   \\
{\em Solution method:}
  %Describe the method solution here.
  To find the bubble-wall profile, the program assumes that tunneling happens along a fixed path in field space. This reduces the equations of motion to one dimension, which can then be solved using the overshoot/undershoot method. The path iteratively deforms in the direction opposite the forces perpendicular to the path until the perpendicular forces vanish (or become very small). To find the phase structure, the program finds and integrates the change in a phase's minimum with respect to temperature.
   \\
%{\em Restrictions:}\\
  %Describe any restrictions on the complexity of the problem here.
%   \\
%{\em Unusual features:}\\
  %Describe any unusual features of the program/problem here.
%   \\
%{\em Additional comments:}\\
  %Provide any additional comments here.
%   \\
{\em Running time:}
  Approximately 1 minute for full analysis of the two-dimensional test model on a 2.5 GHz CPU.
   \\
\end{small}

\section{Introduction}
 
Phase transitions driven by scalar fields likely played an important role in the very early evolution of the Universe. In most inflationary models, the dynamics are driven by the evolution of a scalar inflaton field, while at later times electroweak symmetry breaking is thought to be driven by a transition in the Higgs field vacuum expectation value. Electroweak scale physics is currently being probed by the LHC, so the phenomenology of the electroweak phase transition is of particular interest. A strongly first-order electroweak phase transition would have been a source of entropy production in the early Universe (thereby changing the evolution of its scale with respect to temperature) and produced a stochastic background of gravitational radiation\cite{Witten:1984rs}, perhaps observable by future space-based gravitational radiation observatories\cite{decadalSurvey:2010}. In addition, a strongly first-order electroweak phase transition may have satisfied the Sakharov conditions\cite{Sakharov:1967dj} and been responsible for the current baryon asymmetry of the universe (for recent studies see e.g. Refs.~\cite{Carena:1996wj, Huber:2001xf, Kang:2004pp, Carena:2008vj, Funakubo:2009eg, Chiang:2009fs, Chung:2010cd, Profumo:2007wc, Ham:2010ha, Lee:2004we, Chung:2008aya, Cirigliano:2009yd, Chung:2009qs}),
or may have affected the relic density of (for example) dark matter particles \cite{Wainwright:2009mq, Chung:2011hv, Chung:2011it}.

In the standard model, the electroweak phase transition is not strongly first-order unless the Higgs mass is below $\sim 70$ GeV\cite{Kajantie:1996mn, Csikor:1998eu, Aoki:1999fi}, which is excluded by the current LEP bound of 114.4 GeV\cite{Barate:2003sz}. However, electroweak baryogensis can be saved in extensions to the standard model, many of which include extra dynamic scalar fields (such as two-Higgs-doublet models \cite{Turok:1990zg, Funakubo:1993jg, Davies:1994id, Cline:1995dg, Laine:2000rm, Fromme:2006cm, Cline:2011mm}). The amount of produced baryon asymmetry depends crucially upon the dynamics of the phase transition, and particularly upon the bubble-wall profiles that separate the high- and low-temperature phases. These profiles are fairly easy to calculate if there is only one scalar field, but multiple fields greatly increase the computational complexity.
 
\begin{figure}[h]
   \centering
   \includegraphics[scale=.9]{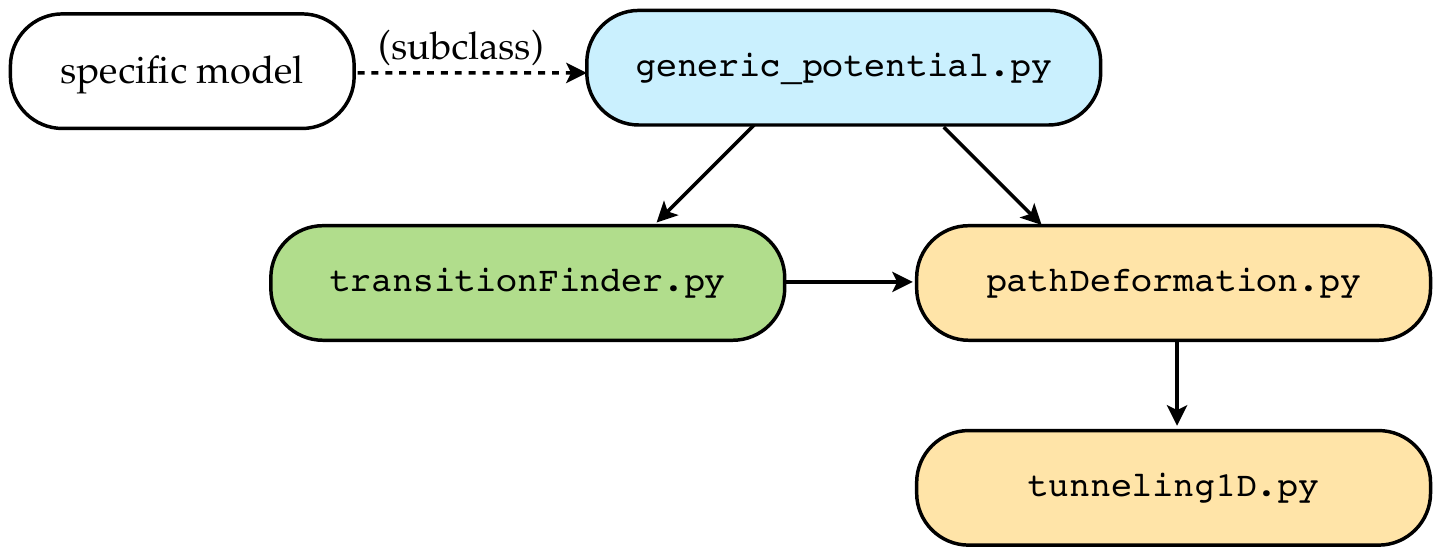} 
   \caption{Overview of the \mycode\ package file structure. The files \texttt{pathDeformation.py} and \texttt{tunneling1D.py} find critical bubble profiles, \texttt{transitionFinder.py} finds the minima of finite temperature potentials as a function of temperature and analyzes phase transitions, and \texttt{generic\_potential.py} defines an abstract class that can easily be subclassed to examine a specific model.}
   \label{fig:overview}
\end{figure}

In this paper, I present an easy-to-use numerical package (\mycode) to analyze phase transitions in finite temperature field theory with multiple scalar fields. The program consists of three basic parts (see fig.~\ref{fig:overview}): modules for finding the tunneling solution (bubble wall profile) between different vacua, a module for finding critical temperatures and phase transitions, and an abstract class to define specific field-theoretic models. In section~\ref{sec:tunneling}, I describe the algorithms for finding bubble wall profiles for both single and multiple fields. Section~\ref{sec:explore} describes the algorithm for finding phase transitions, while section~\ref{sec:structure} describes how one can implement a specific model in a simple program. Finally, I present numerical results in section~\ref{sec:results} and conclude in section~\ref{sec:conclusion}.

To download the latest version of \mycode, visit \url{http://chasm.ucsc.edu/cosmotransitions}.

\section{Calculating bubble profiles}
\label{sec:tunneling}
First-order cosmological phase transitions proceed by the nucleation of bubbles of true vacuum out of metastable false vacuum states. The bubbles have both surface tension and internal pressure, so that large bubbles tend to expand and small bubbles tend to collapse. Critical bubbles---bubbles that are just large enough to avoid collapse---will drive the phase transition.

Given a Lagrangian
\begin{equation}
\label{eq:lagrangian}
\mathcal{L} = \frac{1}{2}(\partial_\mu \vec{\phi}) (\partial^\mu \vec{\phi}) - V(\vec{\phi}),
\end{equation}
where $\vec{\phi}$ is a vector of scalar fields, a critical bubble can be found by extremizing the Euclidean action
\begin{equation}
\label{eq:action}
S_E = \int{d^d x \left[ \frac{1}{2}\left(\partial_\mu \vec{\phi}\right)^2 + V(\vec{\phi})\right]},
\end{equation}
where $d = 4$ (3) for tunneling at zero (finite) temperature. This quantity is critical for finding the nucleation rate, and thereby determining the phase transition temperature and whether or not the transition actually happens (see Refs.~\cite{Coleman:1977py,Callan:1977pt,Linde:1980tt} for seminal work on phase transitions in quantum field theory).
The bubble nucleation rate per unit volume is $\Gamma/V = Ae^{-S_E}$ at zero temperature and $\Gamma/V = Ae^{-S_E/T}$ at finite temperature. The prefactor $A$ is quite difficult to calculate, but it has only weak temperature dependence and can generally be estimated on dimensional grounds (see, e.g., Ref.~\cite{Quiros:1999jp}). By requiring that the expectation value for one bubble to nucleate per Hubble volume is $\sim\mathcal{O}(1)$, one can show that the bubble nucleation temperature for weak-scale fields is given by $S_E/T \sim 140$.

Assuming spherical symmetry, the bubble's equations of motion are
\begin{equation}
\label{eq:eom}
\frac{d^2 \vec{\phi}}{d\rho^2} + \frac{\alpha}{\rho}\frac{d \vec{\phi}}{d\rho} = \nabla V(\vec{\phi}).
\end{equation}
At finite temperature, $\rho$ is simply the spatial radial coordinate and $\alpha=2$. At zero temperature, $\rho^2 = r^2 - t^2$ and $\alpha = 3$. Let $\vec{\phi}_T$ and $\vec{\phi}_F$ denote the true and false vacua, respectively. Then in order for the solution to match the field at infinity, we require that $\vec{\phi}(\infty) = \vec{\phi}_F$. We also demand that $\left.\frac{d\vec{\phi}}{dr}\right|_{\rho=0}=0$ so that bubble is non-singular at the origin.

\subsection{One-dimensional solution}
\begin{figure}[t]
   \centering
   \includegraphics[width=3in]{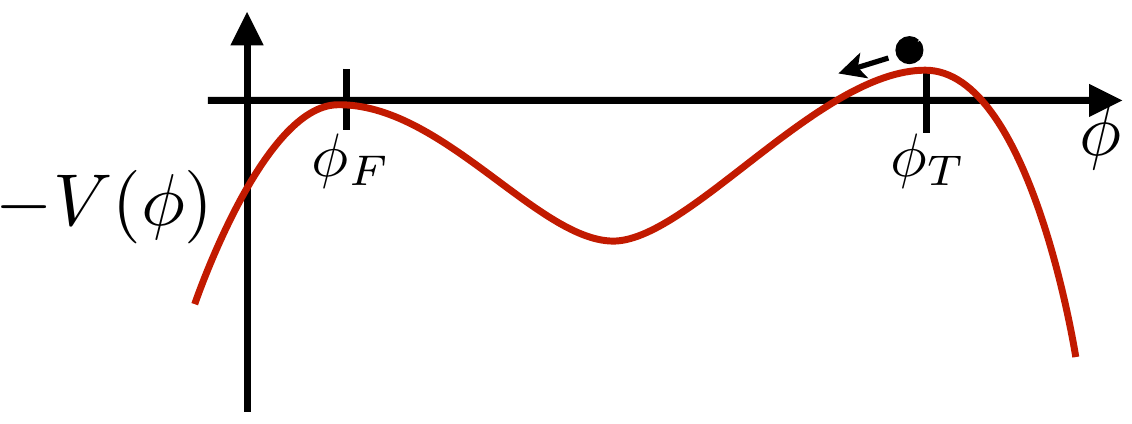} 
   \caption{The equations of motion for a field with a potential $V(\phi)$ can be thought of as the equations for a particle moving in an inverted potential $-V(\phi)$.}
   \label{fig:invpot}
\end{figure}

When the field has only one dimension, the solution to the critical bubble profile can easily be solved by the overshoot/undershoot method (see, e.g., Ref.~\cite{Apreda:2001us}). Here, it is easiest to think of the problem as that of a classical particle moving under the influence of the inverted potential $-V(\phi)$ plus a peculiar looking friction term, where $\phi$ takes on the role of a spatial coordinate and $\rho$ acts as the time coordinate. The problem then is to find the initial placement of the particle near $\phi_T$ such that it rolls down the potential and comes to a stop at $\phi_F$ when $\rho=\infty$ (see figure~\ref{fig:invpot}). If the particle rolls past (overshoots) $\phi_F$, then the initial placement was too close to $\phi_T$. If it doesn't have enough energy to make it to $\phi_F$ (an undershoot), then the initial placement needs to be closer to $\vec{\phi}_T$. Through trial and error, one can find the initial placement to arbitrary precision.

The code presented with this paper follows the general overshoot/undershoot method, implemented by the class \texttt{tunneling1D.bubbleProfile}. Calculation of thick-walled bubbles is straightforward, but thin-walled bubbles require extra consideration. In thin-walled bubbles, the transition from $\phi \approx \phi_T$ to $\phi \approx \phi_F$ happens over a distance much shorter than the bubble's overall size. In the particle analogy, the particle sits very close to $\phi_T$ for a very long time before quickly rolling down the potential and stopping at $\phi_F$. Extreme accuracy in $\phi_0 = \phi(\rho=0)$ would be needed to reliably calculate the wall profile, since a small change in $\phi_0$ would lead to a large change in the radius of the bubble. Instead, I define a new variable $x$ such that $\phi_0 = \phi_T + e^{-x}(\phi_F-\phi_T)$, and use it as the initial condition to vary instead. 

For small $\rho$ and $\phi \approx \phi_T$, the equation of motion can be approximated as
\begin{equation}
\frac{d^2{\phi}}{d\rho^2} + \frac{\alpha}{\rho}\frac{d{\phi}}{d\rho} = \left.\frac{d^2 V}{d\phi^2}\right|_{\phi=\phi_T} (\phi-\phi_T)
\end{equation}
which has the exact solution
\begin{equation}
\phi(\rho) - \phi_T = C \rho^{-\nu} I_\nu \left( \rho \sqrt{b} \right)
\end{equation}
where $\nu = \frac{\alpha-1}{2}$, $b = \left.\frac{d^2 V}{d\phi^2}\right|_{\phi=\phi_T}$, and $I_\nu$ is the modified Bessel function of the first kind. For $\alpha = 2$, this simplifies to $\phi(\rho)-\phi_T \propto \frac{1}{\rho}\sinh( \rho \sqrt{b})$. These can be numerically inverted to find $\rho(\phi)$, which allows one to calculate the approximate radius of a thin-walled bubble without performing any integration. The integration can then start at the edge of the bubble wall, which both increases accuracy and decreases computation time.

\begin{figure}[t]
   \centering
   \includegraphics[width=3in]{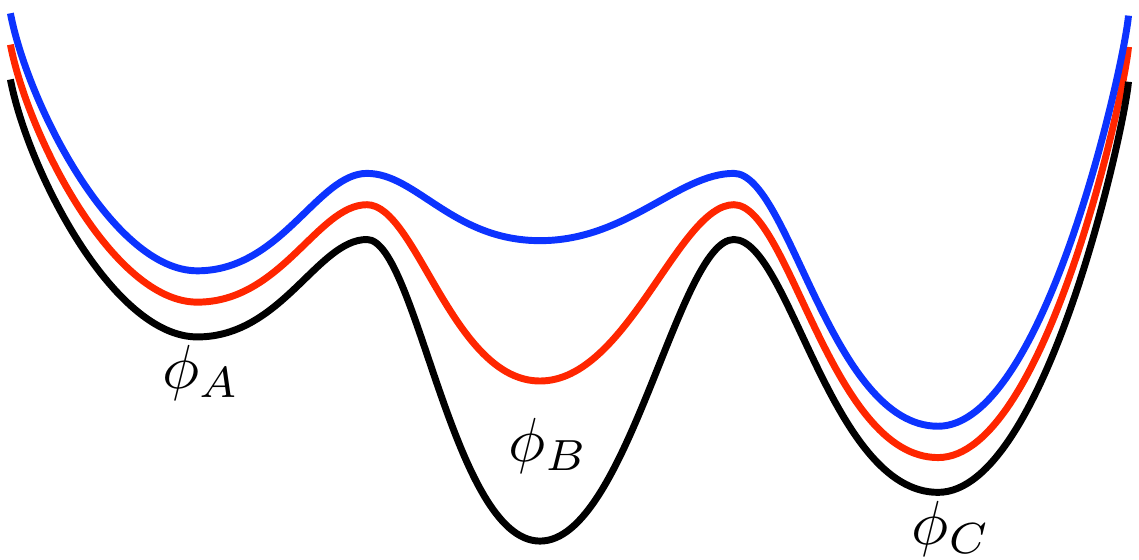} 
   \caption{A problematic potential with more than two minima. A tunneling solution from $\phi_A$ to $\phi_C$ is only guaranteed to exist for the topmost potential (blue line).}
   \label{fig:probV}
\end{figure}

The overshoot/undershoot implementation is fairly robust, but there are situations in which it either does not work or is unreliable. Consider the potentials shown in figure~\ref{fig:probV}. They each have three minima $\phi_A < \phi_B < \phi_C$, with $V(\phi_A) > V(\phi_C)$. If one tries to calculate the bubble profile for tunneling from $\phi_A$ to $\phi_C$, only the topmost potential is unproblematic. There, there will be a solution where the field starts near $\phi_C$, rolls over the bump at $\phi_B$, and ends up at $\phi_A$. In the bottommost potential, no such solution exists. The minimum at $\phi_B$ is the true vacuum of the theory, so the field will tunnel from $\phi_A$ to $\phi_B$, ignoring $\phi_C$ completely. The overshoot/undershoot method will return this solution (any initial $\phi_0 > \phi_B$ will register as an undershoot), but it won't perform any optimization for thin-walled bubbles and may have poor accuracy. In the middle case, with $V(\phi_A) > V(\phi_B) > V(\phi_C)$, there may or may not be a direct tunneling solution from $\phi_A$ to $\phi_C$. Using the particle analogy, it could be that any particle that starts near $\phi_C$ and has enough momentum to get beyond $\phi_B$ will necessarily overshoot $\phi_A$. In this case, the field can only tunnel to $\phi_C$ in two steps: first by tunneling to $\phi_B$, and then by tunneling to $\phi_C$ in a separate nucleation event. The algorithm in \texttt{tunneling1D.bubbleProfile} will only ever return the first of these transitions, and it may have poor accuracy.

\subsection{Multi-dimensional solution and path deformation}

With the solution to the one-dimensional problem in hand, we are ready to tackle the more challenging multi-dimensional problem. The overshoot/undershoot method no longer works because we required the unique topology of the one-dimensional case to determine whether a particular solution overshot or undershot the boundary condition at $\rho=\infty$. Instead, I propose a method of path deformation.

First, assume as an initial guess that the tunneling occurs on some fixed path in field space. That is, $\vec{\phi}_{guess} = \vec{\phi}(x)$, where $x$ parametrizes the path and for simplicity we require that $\left|\frac{d\vec{\phi}}{dx}\right| = 1$. The equations of motion split into two parts---one parallel and one perpendicular to the path:
\begin{eqnarray}
\label{eq:eom_par}
\frac{d^2x}{d\rho^2} + \frac{\alpha}{\rho}\frac{dx}{d\rho} = &\frac{\partial}{\partial x}V[\vec{\phi}(x)] \\ 
\label{eq:eom_perp}
\frac{d^2\vec{\phi}}{dx^2}\left(\frac{dx}{d\rho}\right)^2 = & \nabla_\perp V(\vec{\phi}),
\end{eqnarray}
where $\nabla_\perp V$ represents the components of the gradient of $V$ that are perpendicular to the path. Equation~\ref{eq:eom_par} is the same as the one-dimensional equation of motion, which we can solve by the overshoot/undershoot method. If the path guess $\vec{\phi}(x)$ is correct, then the solution to equation~\ref{eq:eom_par} will also solve equation~\ref{eq:eom_perp}. The trick then is to find the right path. A similar approach of path deformation has been proposed by Ref.~\cite{Cline:1999wi}, while Refs.~\cite{John:1998ip, Konstandin:2006nd, Park:2010rh} use alternate methods to solve the equations of motion. I have previously employed the basic algorithm described below in the context of zero-temperature phase temperatures and singlet scalar dark matter models~\cite{Profumo:2010kp}.

\begin{figure}[t]
   \centering
   \includegraphics[width=3in]{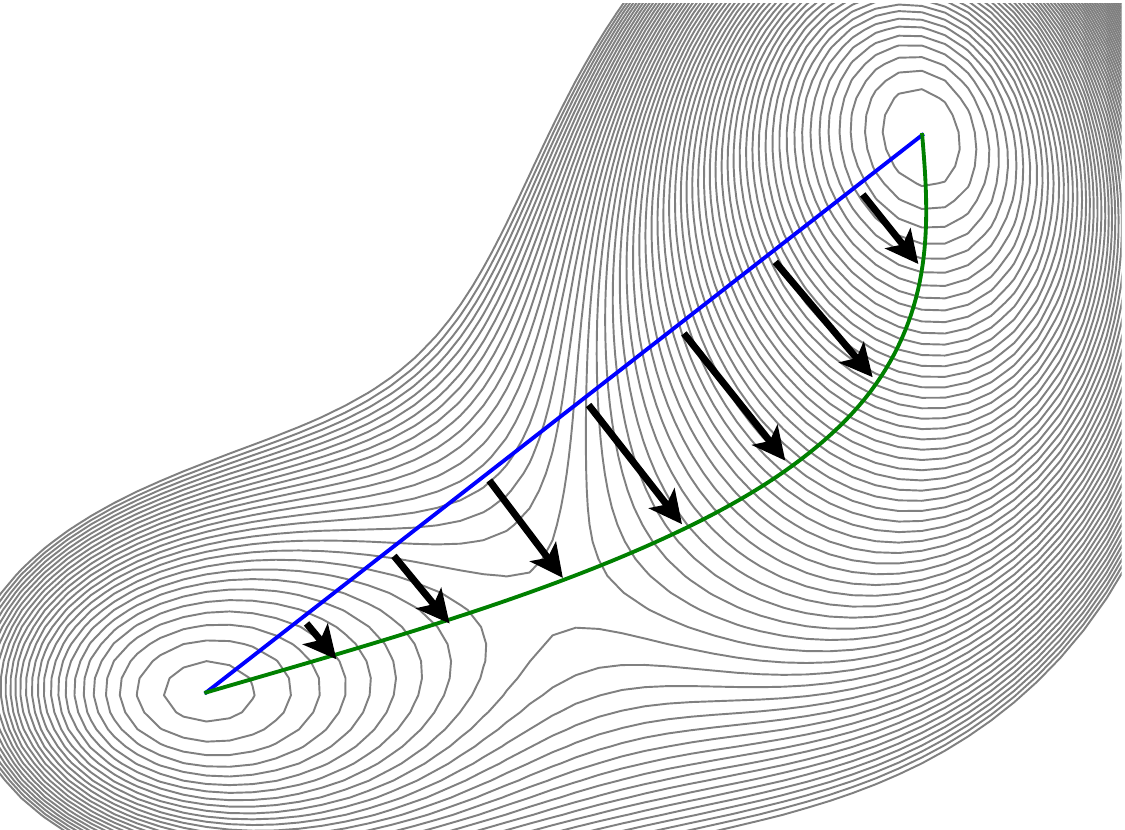} 
   \caption{Path deformation in two dimensions. The normal force exerted on the starting path (blue straight line) pushes it in the direction of the true tunneling solution (green curved line).}
   \label{fig:deform1}
\end{figure}

Using the particle analogy, we can think of the path guess as a fixed track on which the particle moves through the multi-dimensional space. Equation~\ref{eq:eom_par} describes the forces parallel to the track and thus determine the particle's speed. Equation~\ref{eq:eom_perp} does not effect the motion of the particle, but instead determines the normal force $N$ that the track must exert on the particle to keep it from falling off: $N = \frac{d^2\vec{\phi}}{dx^2}\left(\frac{dx}{d\rho}\right)^2 -  \nabla_\perp V(\vec{\phi})$. For the right path, $N=0$. Given a starting guess, one can deform the path to the correct solution by continually pushing it in the direction of $N$ (see figure~\ref{fig:deform1}).

I implement this general method in the \mycode\ package using B-splines in the module \texttt{pathDeformation.py}. Each path is written as a linear combination of spline basis functions plus a linear component connecting its ends: $\vec\phi(y) = \sum_i \beta_i \vec{\phi}_i(y) + (\vec{\phi}_0-\vec{\phi}_F)y + \vec{\phi}_F$, where $y$ parametrizes the path ($0\leq y\leq 1$, and generally $\left|\frac{d\vec{\phi}}{dy}\right| \neq 1$), and $\vec{\phi}_i(0) = \vec{\phi}_i(1) = 0$. 
This fixes the path's endpoints at $\vec{\phi}_0 \sim \vec{\phi}(\rho\!=\!0)$ and $\vec{\phi}_F$. Generally, only a small number of basis functions are needed to accurately model the path ($\sim 10$ per field direction) unless it contains sharp bends or many different curves.

Before any deformation, the algorithm first calculates the bubble profile along the starting path using the overshoot/undershoot method. Then it deforms the path in a series of steps without recalculating either the one-dimensional profile or $\left|\frac{d\vec{\phi}}{d\rho}\right|$. At each step it calculates the normal force for a relatively large number of points ($\sim 100$) along the path, rescales the normal force by $|\vec{\phi}_T-\vec{\phi}_F|/|\nabla V|_{max}$, and moves the points in that direction times some small stepsize.
If the one-dimensional solution is thick-walled (that is, $\vec\phi(\rho\!=\!0)$ is not very close $\vec{\phi}_T$), the algorithm also moves $\vec{\phi}_0$ in the direction of $N$ averaged over the first several points (note that $N(\rho\!=\!0)=0$ as long as the path aligns with $\nabla V$). Otherwise, $\vec{\phi}_0$ stays fixed at $\vec{\phi}_T$. It then recalculates the spline coefficients by a least-squares fit to the moved points with the restriction that the path aligns with $\nabla V$ at $\vec{\phi}_0$ when thick-walled.
The deformation converges when the normal force is much smaller than forces parallel to the path. At this point, the algorithm recalculates $\left|\frac{d\vec{\phi}}{d\rho}\right|$ and deforms the path again. After repeating this procedure a few times, the deformation should converge after a single step and the algorithm will return the final tunneling solution.

\begin{figure}[t]
   \centering
   \includegraphics[width=3in]{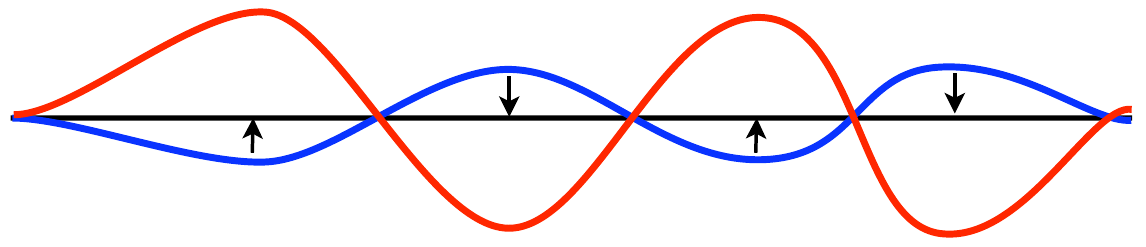} 
   \caption{Example of error correction in deformation. The normal force will push wiggles down towards the straight line solution. But if the stepsize is too large, the wiggles will get reversed and amplified instead.}
   \label{fig:wiggle1}
\end{figure}

Choosing an appropriate stepsize is important. Errors in individual steps are generally self-correcting, but only for small stepsizes. Consider a situation in which the correct tunneling path is a straight line, but we introduce errors in the deformation to add wiggles (see figure~\ref{fig:wiggle1}).
A small stepsize in deformation will tend to smooth out the errors, but a large stepsize will instead reverse and amplify them. By checking for such reversals, the algorithm can keep the stepsize at the appropriate level.

\section{Exploring phase structures}
\label{sec:explore}

In order to determine the characteristics of a phase transition, we must first find where, and at what temperatures, the various phases exist. In theories with spontaneously broken symmetries there is at least one zero-temperature phase and there is generally one high-temperature symmetry-restoring phase. If these phases coexistence at some temperature, then there is likely a first-order phase transition between them. However, even in relatively simple models there can be intermediate phases (see, e.g., Ref.~\cite{Wainwright:2011qy}) which can lead to secondary phase transitions, or can change the quality of the primary transition. Therefore, it is helpful to find the location of the minima as a function of temperature.

By writing the potential about any point $\vec{\phi}'$ as 
\begin{equation}
V(\vec{\phi},T) =  a_i + b_i(\phi_i-\phi_i') + \frac{1}{2}M_{ij}(\phi_i-\phi_i') (\phi_j-\phi_j') + \cdots,
\end{equation}
where $b_i = \frac{\partial V}{\partial \phi_i}$ and $M_{ij} = \frac{\partial^2 V}{\partial \phi_i \partial \phi_j}$,
one can show that there is a (nearby) minimum at $\vec{\phi}_{min} = -M^{-1}\vec{b}$. Therefore, the change in the minimum with respect to temperature is
\begin{equation}
\label{eq:dphidT}
\frac{\partial \vec{\phi}_{min}}{\partial T} = -M^{-1}\frac{\partial \vec{b}}{\partial T}.
\end{equation}
Given a minimum at a single temperature, this allows one to find all of the minima of a given phase as a function of $T$. A singular matrix $M$ indicates a rapid change in the minimum caused by either the disappearance of the phase or a second-order phase transition.

Practically, it is much easier to use this equation in conjunction with a minimization routine than it is to integrate it by itself. The algorithm in \texttt{transitionFinder.traceMinimum} uses the Nelder-Mead downhill simplex method \cite{Nelder:1965zz} to find the local minimum at particular temperatures, and then uses equation~\ref{eq:dphidT} to find how the minimum changes. The former acts as an error check on the latter, which allows for an adaptive stepsize in the temperature. Once the phase disappears, the downhill simplex method can search for a new phase and then trace that. In this manner, one can trace the phase structure of the entire theory (assuming that each phase can be found by minimizing the potential at one of the ends of the other phases).

\section{Structure of a simple program}
\label{sec:structure}

There are essentially three parts to a simple program using my code: the tunneling algorithms and phase tracing algorithms described above, and the implementation of a specific model. Much of this last task happens in the \texttt{generic\_potential} class, which must be subclassed to study any particular theory. 

From the point of view of the finite temperature effective potential, the theory is completely determined by the tree-level potential and field-dependent mass spectrum. The \texttt{generic\_potential} class calculates the one-loop corrections from these masses using $\overline{\text{MS}}$ renormalization \cite{Weinberg:1951ss, 'tHooft:1973mm}
\begin{equation}
\label{eq:V1}
V_1(\vec{\phi}) = \pm\frac{1}{64\pi^2}\sum_i{n_i m^4_i(\vec{\phi})\left[ \log \frac{m^2_i(\vec{\phi})}{Q^2} - c_i\right]},
\end{equation}
where $n_i$ and $m_i$ are the numbers of degrees of freedom and the field-dependent masses of each particle species. The quantity $Q$ is the renormalization scale; $c=1/2$ for gauge boson transverse modes and $3/2$ for all other particles; and the upper (lower) sign is for bosons (fermions). The one-loop finite-temperature corrections are
\begin{equation}
V_1(\vec{\phi},T) = \frac{T^4}{2\pi^2} \sum_i n_i J_\mp \left[ \frac{m_i(\vec{\phi})}{T}\right]
\end{equation}
with
\begin{equation}
J_\mp(x) = \pm \int_0^\infty dy\, y^2 \log\left(1\mp e^{-\sqrt{y^2+x^2}}\right).
\end{equation}
The functions $J_\mp(x)$ are implemented in the module \texttt{finiteT.py} using direct integration and cubic interpolation. Of course, one can add additional structure (such as counter-terms) to a subclass.

To create a fully functioning model using the \mycode\ package, one need only subclass \texttt{generic\_potential} and overwrite three functions: the initialization function \texttt{init()} to specify the number of field dimensions, the tree-level potential function \texttt{V0()}, and the mass-spectrum functions \texttt{boson\_massSq()} and \texttt{fermion\_massSq()}. Additionally, one should overwrite \texttt{approxZeroTMin()} to return the approximate locations of the zero temperature minima, especially if more than one minimum exists. Calling the function \texttt{getPhases()} will run routines from \texttt{transitionFinder.py} and calculate the phase structure of the theory. The critical temperature(s) (the temperature of degenerate minima between two phases) can be found by calling the function \texttt{calcTcTrans()}, while the function \texttt{calcFullTrans()} will find the amounts of supercooling and critical bubble profiles for each transition.

\section{Numerical results}
\label{sec:results}

\subsection{Deformation}

To test the path deformation and tunneling routines, I use a simple potential given by
\begin{equation}
\label{eq:testV}
V(x,y) = \left(x^2 +y^2\right) \left[1.8(x-1)^2 + 0.2(y-1)^2 - \delta\right].
\end{equation}
This has one local minimum at $x=y=0$, and a global minimum near $x=y=1$. For $\delta\ll 1$, the phases are nearly degenerate and any tunneling between them will be thin-walled.

\begin{figure}[t]
   \centering
   \includegraphics[scale=.8]{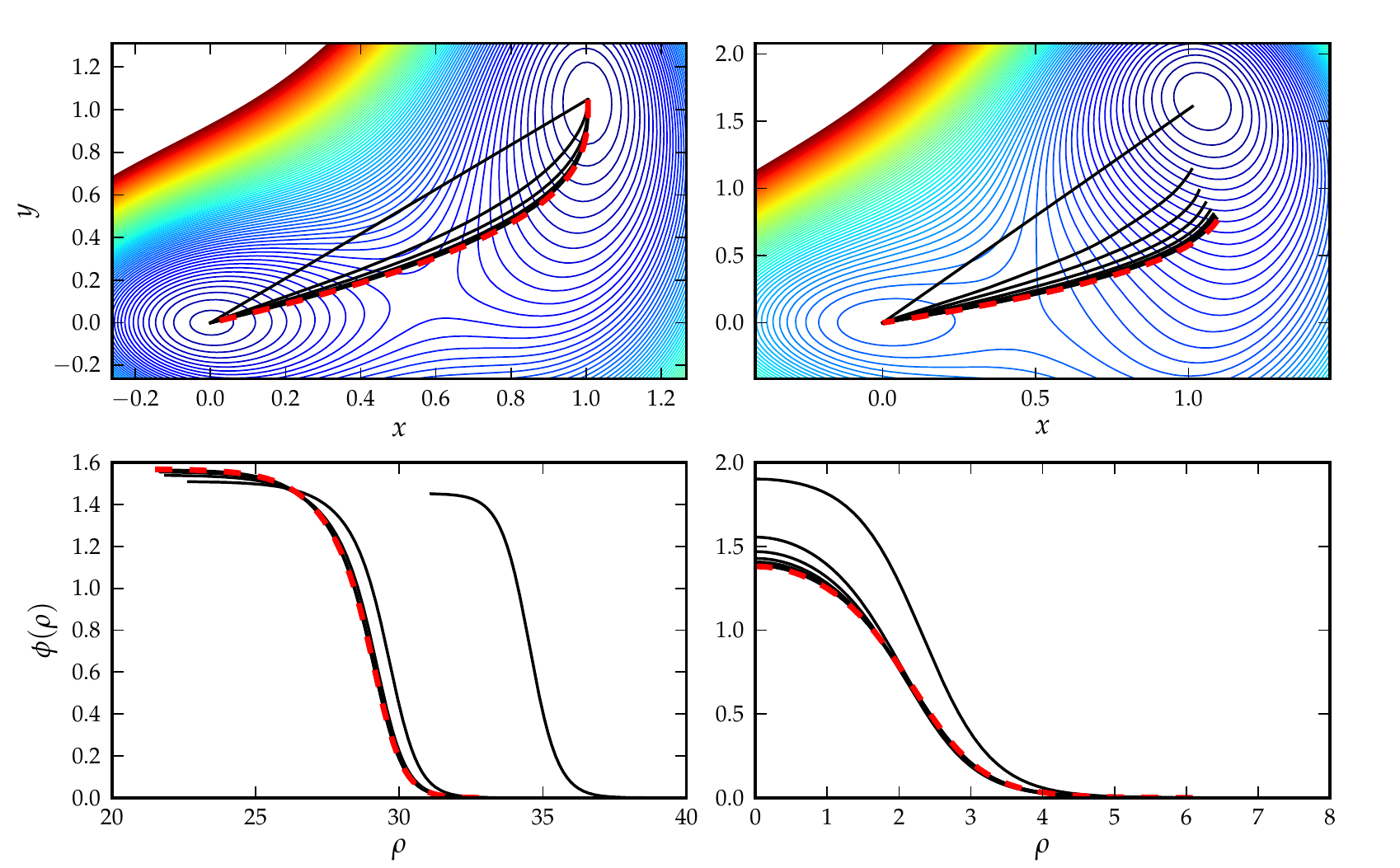} 
   \caption{Top: the potential $V(x,y)$ given by equation~\ref{eq:testV} for $\delta = 0.02$ (left) and $\delta=0.4$ (right). Black lines show successive deformations of the tunneling path, while red dashed lines show direct integration of equation~\ref{eq:eom} with manually chosen initial conditions. Note that the $x$ and $y$ axes are not to scale, so $\nabla V$ does not appear perpendicular to the contour lines. Bottom: the bubble profiles associated with each of the above deformations, where $\phi(\rho)$ is measured along the path.}
   \label{fig:testDeformation}
\end{figure}

I run the \texttt{pathDeformation.fullTunneling} class for both thin ($\delta = 0.02$) and thick-walled ($\delta = 0.4$) potentials, with results shown in figure~\ref{fig:testDeformation}. Each line represents 15 individual deformation steps with adaptive stepsizes on the order of 0.005. In this case, and in general, the thick-walled case converges more slowly due to the added complication of moving $\phi_0=\phi(\rho\!=\!0)$ with each step. The thin-walled case converges in $\sim 60$ steps, while the thick-walled case takes $\sim 150$ steps.

To check the deformation solution, I numerically integrate the equations of motion~\ref{eq:eom}, using manual trial and error to find the correct initial conditions. These are shown in figure~\ref{fig:testDeformation} as the red dashed lines. The deformation algorithm gets extremely close to---but not exactly to---the integrated solution. In the thin-walled case, the deformed path is within 0.1\% of the integrated solution, while in the thick-walled case the error is 0.3\%.

\subsection{Calculating transition temperatures}

To demonstrate the transition finding algorithms, I consider a theory with two scalar fields and a tree-level potential
\begin{equation}
V_0(s_1, s_2) = \frac{1}{8}\frac{m_1^2}{v^2} \left(s_1^2 - v^2\right)^2 + \frac{1}{8}\frac{m_2^2}{v^2}^2 \left(s_2^2 - v^2\right) - \mu^2 s_1 s_2. 
\end{equation}
When $\mu^2 = 0$, the theory has four identical minima located at $(s_1,s_2) = (\pm v, \pm v)$ and $(s_1,s_2) = (\pm v, \mp v)$. The effect of $\mu^2 >0$ is to lower the minima at $ (\pm v, \pm v)$ and raise those at  $(\pm v, \mp v)$. The tree-level scalar mass-matrix is
\begin{equation}
m^2_{ij}(s_1,s_2) = \frac{1}{2 v^2}\begin{pmatrix} {m_1^2 (3s_1^2 - v^2)} & {-\mu^2} \\ {-\mu^2} & {m_2^2 (3s_2^2 - v^2)} \\ \end{pmatrix},
\end{equation}
so that the masses at the tree-level minima are $m_1$ and $m_2$ when $\mu^2 = 0$. I also add an extra bosonic degree of freedom with the field-dependent mass
\begin{equation}
m_X(s_1, s_2) = y_A^2 \left(s_1^2+s_2^2\right) + y_B^2 s_1 s_2.
\end{equation}
The peculiar coupling $y_B$ is designed to lift the minima at $(\pm v, \pm v)$ relative to $(\pm v, \mp v)$ at finite temperature so that there may be a phase transition between the two.

I examine a model with the parameters given in table~\ref{tab:params}, where $Q$ is the renormalization scale used in equation~\ref{eq:V1} and $n_X$ is the number of degrees of freedom assigned to the extra boson with mass $m_X$. For this particular choice of parameters all tree-level minima are (meta)stable, but loop corrections destroy the zero-temperature minima at $(\pm v, \mp v)$. I chose the parameters primarily to showcase multiple transitions within a single model, not for any physically motived purpose. The tunneling condition is $S_E/T = 140$.

\begin{table}[t]
\caption{Model parameters}
\begin{center}
\begin{tabular}{|c|c|c|c|c|c|c|c|}
$m_1$ & $m_2$ & $\mu$ & $v$ & $Q$ & $y_A^2$ & $y_B^2$ & $n_X$\\ \hline
120 GeV & 50 GeV & 25 GeV & 246 GeV & 246 GeV & 0.10 & 0.15 & 30 \\
\end{tabular}
\end{center}
\label{tab:params}
\end{table}%

\begin{figure}[p]
   \centering
   \includegraphics[scale=.8]{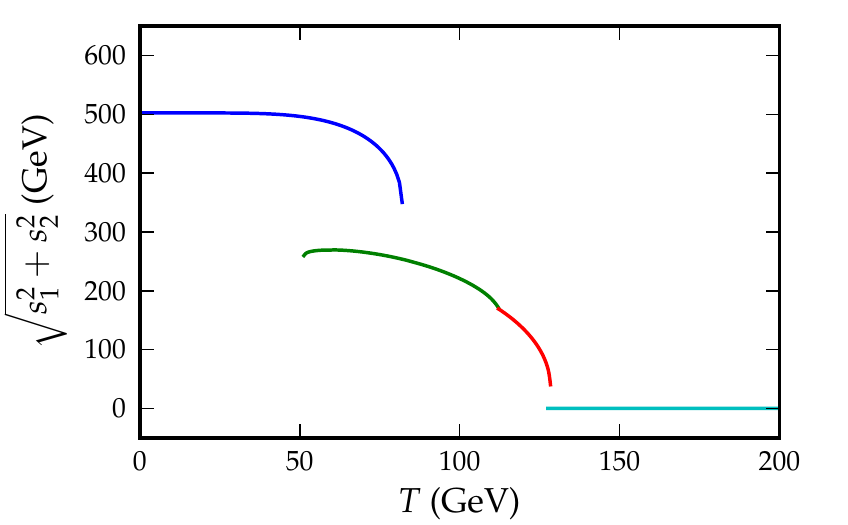} 
   \caption{The position of the minima as a function of temperature.}
   \label{fig:phases}
\end{figure}

\begin{figure}[p]
   \centering
   \includegraphics[scale=.8]{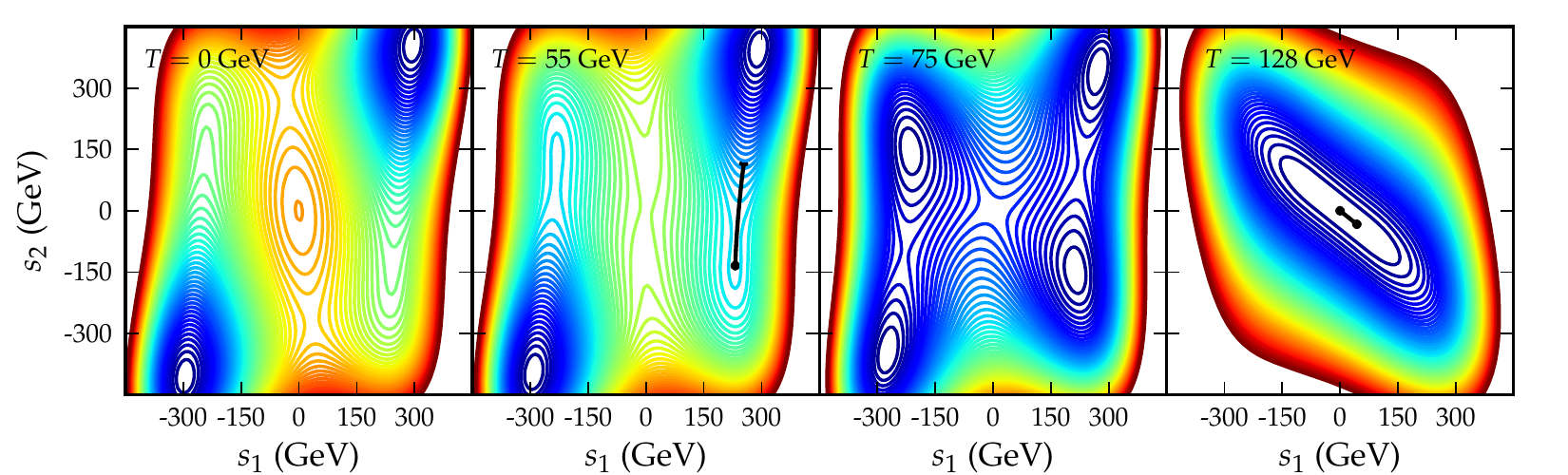} 
   \caption{The finite-temperature effective potential $V(s_1,s_2,T)$ at $T=0$, 55, 75, and 128 GeV. Thick black lines at $T = 55$ and 128 GeV show the tunneling paths for the phase transitions.}
   \label{fig:finiteTV}
\end{figure}

\begin{figure}[p]
   \centering
   \includegraphics[scale=.8]{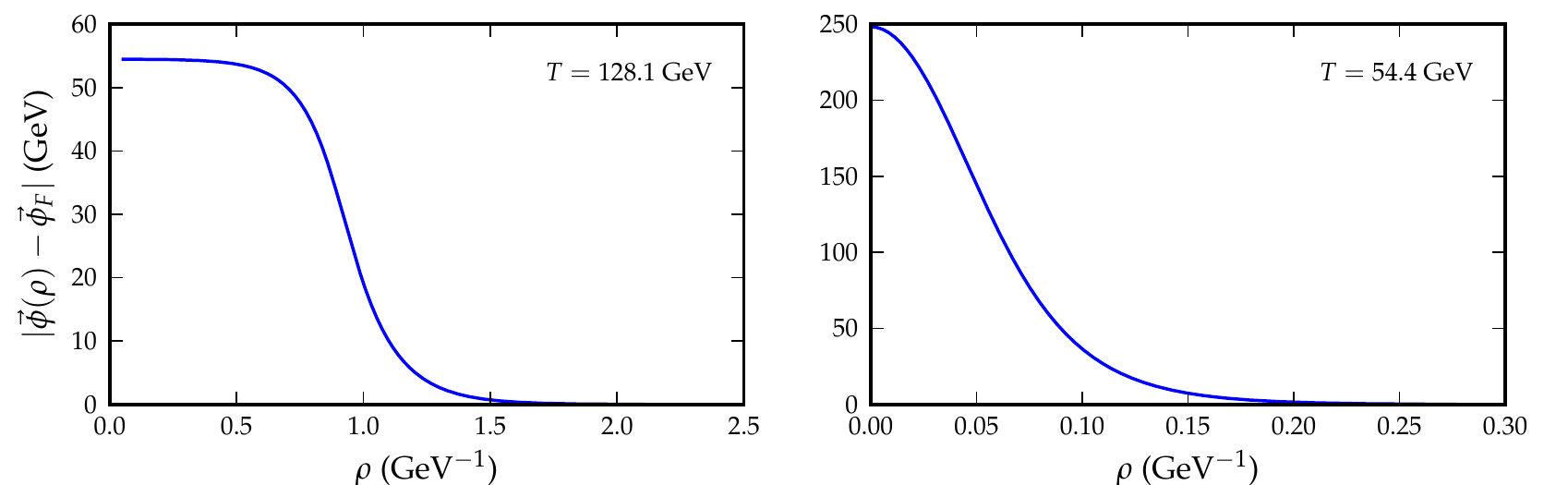} 
   \caption{Bubble profiles along the tunneling paths for both the high- and low-temperature phase transitions. The high-temperature profile is (relatively) thin-walled, while the low-temperature profile is very much thick-walled.}
   \label{fig:transProfile}
\end{figure}

Figure~\ref{fig:phases} shows the phase structure of the model, and figure~\ref{fig:finiteTV} shows the general evolution of the potential as a function of temperature. At high temperatures, there is a single phase with $s_1=s_2=0$. When the temperature drops to 128.2 GeV, a new phase appears at $(s_1, s_2) = (43, -33)$ GeV. The system tunnels to this phase at $T=128.1$ GeV via a first-order phase transition (see figure~\ref{fig:transProfile}), and the high-temperature phase disappears by $T = 127.6$ GeV. At $T=112.2$ GeV, there is a small discontinuity due to the non-analyticity of $J_-(m/T)$ when $m=0$. At the default resolution the algorithm registers this as a second-order phase transition, but for a small range of temperatures (112.14--112.22 GeV) there are actually two distinct phases separated by $\Delta\phi\approx 2$ GeV. Immediately below this the system can be thought of as being in the phase associated with the tree-level vev at $(s_1, s_2) = (+v, -v)$. As the system cools, the phase associated with $(s_1, s_2) = (+v,+v)$ appears, and by $T=75.2$ GeV the two phases are degenerate. Below this the former phase is metastable, but it does not transition until $T=54.4$ GeV, which is just above the point at which it disappears. Thus, the final tunneling is very thick-walled (see figure~\ref{fig:transProfile}).

In order to reproduce these calculations, one needs to create an instance of the class \texttt{testModels.model1} with parameters given by table~\ref{tab:params} and call the class functions \texttt{getPhases()}, \texttt{calcTcTrans()} and \texttt{calcFullTrans()} with their default parameters. Much of the plotting can be handled by the functions \texttt{plot2d()} and \texttt{plotPhasesPhi()}.

To check that the code works with three scalar fields, I simply add an extra field to the potential such that its minimum is always zero: $V_0(s_1,s_2,s_3) = V_0(s_1,s_2) + v^2 s_3^2$. I then rotate $s_2$ and $s_3$ by $45^\circ$. This gives an easily visualizable potential with non-trivial minima. Running the above commands produces the same results as in the two-dimensional model within the default error tolerances. I successfully tested a four dimensional model in a similar way. 
In four dimensions, the transition finding routine with default resolution labels the highest temperature transition as second-order instead of weakly first-order, but this can be corrected by increasing the resolution.
Higher dimensional models have not been tested, but the code was written to support an arbitrary number of scalar fields.

\section{Conclusion}
\label{sec:conclusion}

I presented the publicly available \mycode\ package to analyze cosmological phase transitions. This included algorithms to find the temperature-dependent phase minima, their critical temperatures, and the actual nucleation temperatures and tunneling profiles of the transitions. I introduced a novel method of path deformation to find the profiles, which I then demonstrated in simple test cases to accuracies of order $\sim 0.1$\%. The deformation algorithm has been successfully tested in 2 and 3 dimensions with both thick- and thin-walled profiles, but it should work in any number of higher dimensions as well.

\mycode\ is designed to be easily extensible, with minimal work needed on the part of the model builder. A new model can be created by subclassing the \texttt{generic\_potential} class and specifying only it's number of dimensions, tree-level potential, and field-dependent particle spectrum. This will (hopefully) allow for the quick analysis of phase transitions in many extensions to the standard model.

\section*{Acknowledgments}
I am grateful to Stefano Profumo for many helpful comments over the course of this project. I receive support from the National Science Foundation.

\bibliographystyle{elsarticle-num}
\bibliography{codePaper}

\end{document}